\documentclass[twocolumn,aps,prl,showpacs]{revtex4}

\usepackage{graphicx}
\usepackage{dcolumn}
\usepackage{color}
\usepackage{bm}
\usepackage{ulem}

\begin{document}

\title{Identifying contact effects in electronic conduction through buckyballs
on silicon}

\bigskip

\author{G-C. Liang }
\author{ A. W. Ghosh}
\address{ School of Electrical and Computer Engineering, Purdue
University, W. Lafayette, IN 47907}%

\medskip

\widetext

\begin{abstract}
We present a theory of current conduction through buckyball
(C$_{60}$) molecules on silicon by coupling a density functional
treatment of the molecular levels embedded in silicon with a
non-equilibrium Green's function (NEGF) treatment of quantum
transport. Several experimental variations in conductance-voltage
(G-V) characteristics are quantitatively accounted for by varying
the detailed molecule-silicon bonding geometries. We identify how
variations in contact surface microstructure influence the number,
positions and shapes of the conductance peaks, while varying
separations of the scanning probe from the molecules influence
their peak amplitudes.
\end{abstract}
\bigskip

\bigskip


\maketitle

Molecular electronics represents an ultimate dream for nanoscale
material and device engineering. Along with well characterized,
reproducible experiments, quantitative models for molecular
conduction are crucial for a proper understanding and benchmarking
of this emerging field, and for the exploration of novel device
paradigms. A persistent problem has been incomplete knowledge of
metal contact microstructures and their influence on conduction.
In this respect, a semiconducting substrate provides a superior
test-bed due to its well studied surface chemistry for transport
measurements \cite{rhersam,rrak}. It is therefore worthwhile to
develop and refine our knowledge base using a familiar molecule
bonded on a well-characterized silicon substrate that leaves
little wiggle room for theory.

Among various molecules probed using scanning tunneling
spectroscopy (STS), buckyball ($C_{60}$) molecules stand out for
their unique well calibrated bandstructure, alkali metal doped
superconductivity, switching and optoelectronic properties
\cite{rDressbook}. Although STS studies of buckyballs on metals
have allowed detailed comparison with theory \cite{rLouie}, they
do not reveal much information about their underlying contact
microstructure. In contrast, buckyballs on silicon exhibit
considerable variation in their G-V characteristics depending on
the nature of their covalent bondings with the surface dimers
\cite{ryao,rdekker,rhersamC60,rbolotov}.

\begin{figure}[ht]
\vspace{2.4 in}
\includegraphics{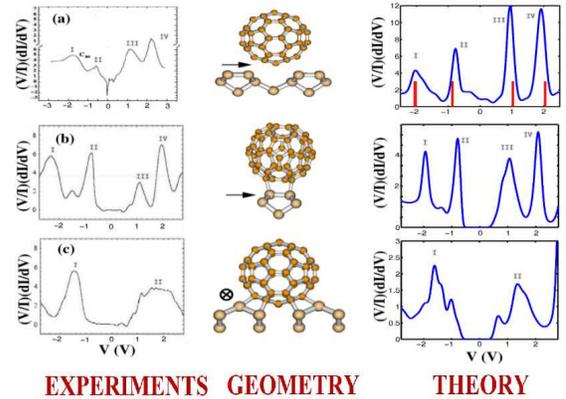}
\caption{Different STS measurements \cite{ryao,rdekker} on
$C_{60}$ molecules docked onto Si(100) 2$\times$1 surface (left
panel). We attribute different bonding geometries (middle panel)
for each experiment, leading to a theoretical G-V (right panel)
that agrees quite well with the corresponding measurement. The
arrow represents the dimer direction of the reconstructed surface
(cross going into the page). The upper geometry corresponds to
$C_{60}$ physisorbed on four surface dimers, the middle one
represents the buckyball chemically bonded with a single surface
dimer, while the bottom one has the molecule lowered into the
trough caused by a missing dimer.} \label{f0}
\end{figure}

In this paper, we explore conduction through buckyballs on
silicon, and correlate observed variations in their G-Vs with
variations in their contact bonding geometries (Fig.~\ref{f0}). A
variation in the nature of the molecule-substrate bonding  leads
to a variation in the number and shapes of conductance peaks,
while a variation in the tip-sample tunneling gap leads to a
variation in the relative peak heights. Our theoretical
formulation thus serves a dual purpose: on one hand, it tests our
quantitative model for molecular conduction, in particular on a
sophisticated semiconducting substrate; on the other hand, it
provides useful insights that allow us to deconstruct the role of
contact geometry on molecular conduction.

\begin{figure}[ht]
\vspace{2 in} \includegraphics{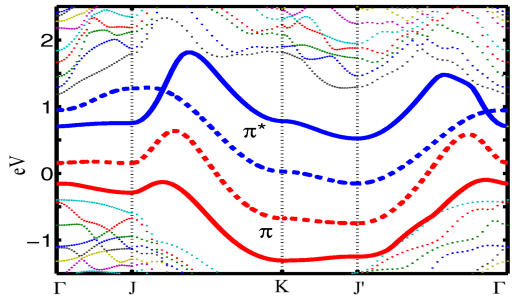} \caption{ Calculated surface bandstructure of
asymmetric dimer (solid lines) and symmetric dimer (dashed lines)
reconstructions along Si(100)-2x1 using EHT bulk Si parameters
\cite{rcerda} and optimized surface geometries \cite{rRamastad}.}
\label{f1}
\end{figure}

{\it{Theoretical technique.}} We calculate molecular conduction by
coupling an electronic structure calculation for the molecule and
the contacts with a treatment of quantum transport using NEGF
\cite{rdatta,rdamle}. The C$_{60}$ structure and Hamiltonian are
obtained using density functional theory within the local density
approximation (LDA). The reconstructed surface geometry of Si(100)
is obtained by LDA optimization with a norm-conserving
pseudopotential in a plane-wave basis \cite{rRamastad}. Although
it is possible in principle to describe the silicon bandstructure
using DFT, as has been customarily done in the past for metal
substrates \cite{rdamle,rguo}, adapting the same process to
semiconductors is quite challenging, given the complicated
bandstructure, extended band-bending and incomplete screening,
reconstruction, and surface chemistry of silicon. Fortunately,
within the NEGF formalism one can formally partition the problem
so that the only quantum effect of the silicon substrate that the
molecule is sensitive to resides in its surface Green's function.
In the past, we developed a technique for combining different
electronic structure codes by matching their interfacial Green's
functions expressed in two different atomistic basis sets. The
match is exact for two equally complete basis sets and a best-fit
otherwise, and only assumes local separability of their
one-electron potentials \cite{rmixed,rrak}.

We use an Extended Huckel (EHT) type model parametrized by Cerda
et. al. \cite{rcerda} to generate a good quantitative description
of the bulk silicon bandstructure (calibrated with LDA+GGA
calculations), and also the surface band structure of 2$\times$1
reconstructed Si-(100) (calibrated with experiments and GGA
calculations \cite{rgw}). To check the properties of Si(100)-2x1
reconstructed surface properly, a slab of 13 silicon layers with a
hydrogen-passivated bottom layer is used to simulate the
bandstructure of Si(100)-2x1 asymmetric dimer (AD) reconstruction
and symmetric dimer (SD) reconstruction. The red and blue solid
lines in Fig~\ref{f1} represent the $\pi$ and $\pi^*$ states of
Si(100)-2x1 AD reconstruction while the dashed lines represent the
corresponding $\pi$ and $\pi^*$ states of the SD reconstruction.
The former clearly shows a bandgap $\sim$ 0.6 eV while the latter
shows a continuum of states in the bulk bandgap region. After
benchmarking these properties (details will be published
elsewhere), the recursive surface green's function is computed for
the semi-infinite silicon substrate.  We then use a mixed-basis
method \cite{rmixed} to transfer the Si(100) surface Green's
function computed in the EHT Slater type orbital (STO) basis into
a 6-31g(d) basis set that is then connected with a DFT/6-31g(d)
Hamiltonian for the molecule.

While the molecule and substrate are modeled atomistically, we
employ a simpler treatment of the STM tip using a self-energy
$\Sigma_2(d_{0})$ \cite{rrak}, where d$_{0}$ is the tip to
molecule bond length. More sophisticated models could be used to
describe tip-sample interactions in experiments with well
characterized tip structures. Vacuum tunneling is described using
a typical WKB factor \cite{rWKB} premultiplying $\Sigma_2(d_0)$,
making the net self-energy $\Sigma_2$ energy and distance
dependant. Our approach includes the bias-dependent barrier
profile and agrees quantitatively with measured STS spectra on
bare silicon.
\begin{figure}
\vspace{2.3in} \includegraphics{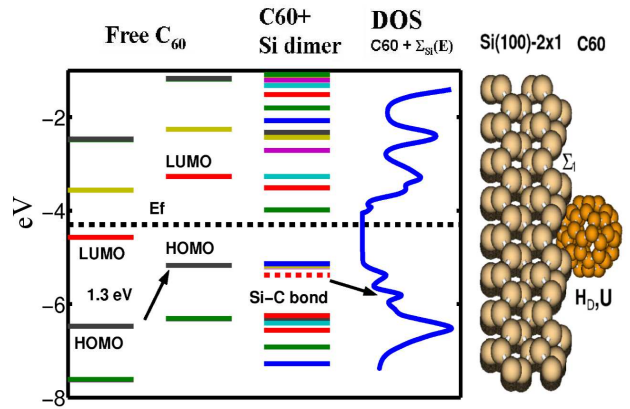} \caption{ Energy level alignment between C$_{60}$ and
silicon. The left panel shows the energy levels of isolated
$C_{60}$, with the calculated Fermi energy of the doped silicon
shown as a dashed line. The C$_{60}$ energy levels are shifted by
1.3 eV due to self-consistent charging driven by the workfunction
difference between the molecule and the substrate, and the
corresponding charge transfer from Si to C$_{60}$. Chemisorption
creates new levels due to SiC bonds, easily seen by adding a
silicon dimer to the molecule and passivating the cluster at the
bottom. Finally, including the silicon self-energy amounts to
adding the entire silicon substrate and generates a continuum
molecular band. The right panel shows the density of states
corresponding to the geometry in the middle of Fig.~\ref{f0}. }
\label{f2}
\end{figure}

\begin{figure*}
\vspace{2.7 in} \includegraphics{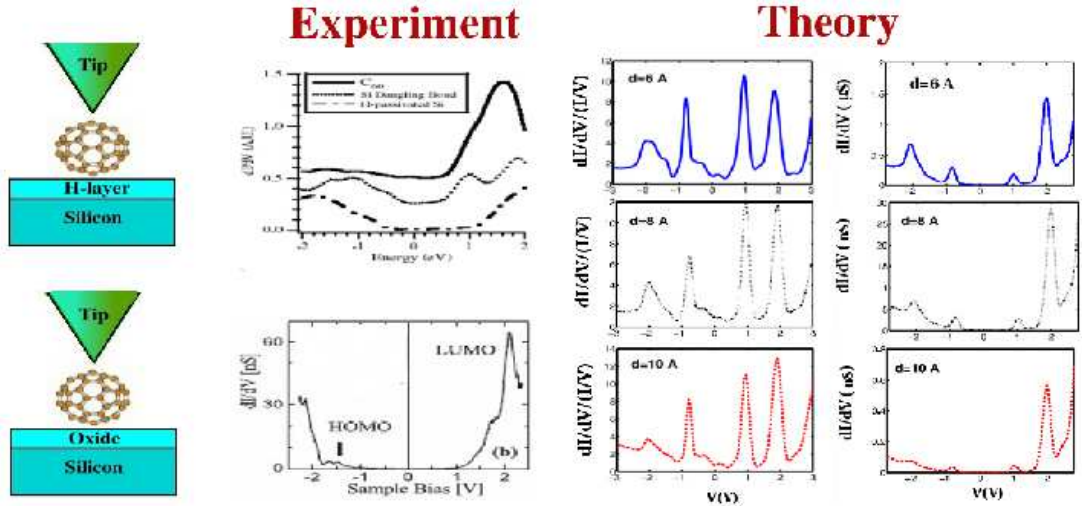} \caption{ Introducing tunnel barriers between the STM
and the sample or the sample and the substrate (far left)
de-emphasizes the HOMO levels relative to the LUMO ones due to
their relatively larger barrier heights, accounting for
measurements (left) with little or no signature of HOMO levels
\cite{rhersamC60,rbolotov}. Although the unnormalized conductance
shows this effect with increasing tip-sample separation $d$
(right), normalizing the conductance (center) as in Fig.~\ref{f0}
restores these peaks.} \label{f3}
\end{figure*}

{\it{Equilibrium band diagram.}} We start by describing the
silicon-buckyball bonding chemistry, band formation, and the
corresponding band alignment due to charge transfer. DFT
(LSDA/6-31g) gives a good description of the energy levels for
isolated C$_{60}$. The highest occupied molecular orbital (HOMO)
is at -6.5 eV while the lowest unoccupied molecular orbital (LUMO)
is at -4.6 eV  relative to vacuum. Once the buckyball connects to
n-doped silicon,  electrons are transferred from Si to C$_{60}$
because the Fermi level of Si is higher than the C$_{60}$ LUMO.
Self-consistent calculations with a Hubbard-type capacitive
charging energy \cite{rsc} yield a net charge transfer that raises
the energy levels of C$_{60}$ by about 1.3 eV, which is very close
to the difference in workfunction between C$_{60}$ and Si. The
charging energy of the Hubbard hamiltonian \cite{hubbard1} is
chosen to be 1.2 eV, consistent with experiments involving
C$_{60}$ on metal and with solid C$_{60}$ surface \cite{hubbard2}.

 Fig.~\ref{f2} explains the energy level diagram using
the geometry in the middle of Fig.~\ref{f0} as an example. The
solid lines represent the energy levels of an isolated buckyball,
while the dashed line represents the fermi level of bulk Si
calculated from the experimental doping levels \cite{ryao}. In
addition to the charge transfer and band-alignment driven by
electrostatics, there is also substantial transfer of spectral
weight from Si to C$_{60}$ leading to the formation of
bonding-antibonding pairs. The energy levels of C$_{60}$ bonded with a
single surface silicon dimer system shows the effect of bonding,
which leads to both level rearrangement and level creation. A
wave-function plot of those levels shows a lot of hybridization
between C and Si. The rightmost panel shows the density of states
of C$_{60}$ with the Si surface which provides the proper boundary
conditions of the open system described with an energy-dependent
self-energy. A clear peak appears between the HOMO and HOMO-1
levels in the STS (i.e., between peaks marked I and II), which we
attribute to the Si-C bond arising from strong Si and C
hybridization.

{\it{Results: Peak positions and heights.}} Fig.~\ref{f0} shows
the calculated G-Vs for C$_{60}$ docked on a clean silicon surface
with different bonding geometries. The conductances are normalized
using an averaging procedure adopted in the experimental analyses
\cite{rChen,rSaridbook}. The upper set of figures corresponds to
C$_{60}$ physisorbed on the Si (100)-2$\times$1 surface. The
bottom of the buckyball is kept 2.1 $\AA$ away from the Si surface
dimer to ensure weak coupling. Under these circumstances, the STS
probes the bare C$_{60}$ electronic structure with the molecular
levels (Fig.~\ref{f2}) generating G-V peaks marked I , II, III,
and IV in Fig.~\ref{f0}. The vertical bars in Fig.~\ref{f0} denote
the LDA/4-31g HOMO-1, HOMO, LUMO, and LUMO+1 levels of isolated
C$_{60}$ rigidly shifted by 1.2 eV due to charging as before,
although the precise shift differs due to the different Fermi
energies of the n and p-Si substrates. The excellent
correspondence between the isolated C$_{60}$ levels and the
conductance peaks thus gives us an elementary interpretation of
the STS data in the upper geometry of Fig.~\ref{f0}.

The bonding geometry between C$_{60}$ and the Si(100) surface
changes upon annealing from physisorption to chemisorption
\cite{rChenC60}, corresponding to the middle and the bottom sets
of plots in Fig.~\ref{f0}. We consider two prominent chemisorption
geometries based on experimental suggestions. The first consists
of C$_{60}$ chemically bonded with a Si surface dimer that
straddles diametrically opposite ends of a C$_{60}$ hexagon. The
experimental dI/dV/(I/V)s measured using STS are reproduced by our
density functional conductance calculation, with the only variable
being the geometry itself. The four main marked peaks still arise
from the isolated C$_{60}$ energy levels discussed above. In
addition, our calculation reveals an extra smaller peak between
peaks I and II, as in the experiments. This small peak has also
been observed by ultraviolet photoemission spectroscopy (UPS)
measurements on the C$_{60}$/Si(100)-2$\times$1 system
\cite{rups}. The origin of this secondary peak is from strong
covalent Si-C bonding, seen in the correlation diagram between
C$_{60}$ and a surface Si dimer in Fig.~\ref{f2} and also in the
corresponding density of states.

The bottom plot of Fig.~\ref{f0} shows the experimental and
calculated conductances corresponding to a different chemisorption
geometry realized upon annealing, consisting of a missing silicon
dimer that causes the buckyball to drop into the empty trough
\cite{ryao}. It is clear from the bonding geometry (middle panel),
that the closer proximity with the surface leads to the
establishment of more covalent bonds, radically altering the
electronic structure of bare C$_{60}$. The STS G-V in (c) is
qualitatively different from the geometries in (a) and (b). There
is one clear HOMO peak for negative substrate bias and a broadened
LUMO level in the positive direction replacing the four original
peaks. Our simulation captures the main features of this
experiment. The original Si-C peak goes up to become the main
negative bias peak, while several additional Si-C peaks are formed
near the conducting LUMO levels due to the additional bondings. In
addition, our calculation generates spurious peaks from the
unrelaxed C$_{60}$ structure adopted in our calculation for
convenience. We believe that strong chemisorption would deform the
C$_{60}$ near the bottom, eliminating these extra peaks by
bonding-antibonding splitting. We leave a detailed study of
C$_{60}$ relaxation for future work.

In addition to the number and positions of the peaks, other
experiments show variations in the conductance peak heights
\cite{rdekker,rbolotov,rhersamC60} (Fig.~\ref{f3}). A possible
origin is the differing tip-sample spacings in these experiments.
A WKB treatment of tunneling through varying vacuum thicknesses
provides a qualitative explanation. Using the physisorbed geometry
in Fig.~\ref{f0}(a) as an example, we find that increasing the
tip-sample gap from 6 $\AA$ to 8 $\AA$ and then to 1 nm
progressively deemphasizes the role of the HOMO levels in
comparison to the LUMO levels. Note however that such WKB factors
are eliminated in the earlier log-normalized dI/dV/(I/V) plots
(Fig.~\ref{f3}), as expected \cite{rChen} but show up in the
unnormalized dI/dV conductance plots (Fig.~\ref{f3}). From the
dI/dV vs V in Fig.~\ref{f3}, it is very clear that the thicker
barrier cuts down the HOMO contributions exponentially to within
the noise levels of the experiment. Electrons tunneling from the
HOMO level in the negative bias direction encounter a higher
tunneling barrier than electrons from the STM filling the LUMO
higher up. The WKB approximation suggests that each peak height is
reduced roughly in proportion to $\exp{(-2kd)}$ where $k$
represents the decay constant of the corresponding level and $d$
is the tip-sample separation. We therefore believe that
measurements with no clear HOMO peaks are performed with an STM
tip substantially removed from the molecule. Furthermore, the
experiments in Fig.~\ref{f3} have an extra tunneling barrier
between the molecule and the substrate (a hydrogen passivation
layer in the top example and an oxide layer in the bottom
\cite{reta}). This additional barrier would further de-emphasize
the HOMO contributions, practically eliminating them from the
unnormalized conductance curves.

In summary, we have used C$_{60}$ on silicon to demonstrate our
capacity to theoretically deconstruct the role of contact
microstructure on molecular conduction. The unexplained features
are the peak broadenings, which should depend on coupling with the
dimer and the molecular vibrational modes. A treatment of vibronic
scattering would indeed be worth pursuing both for its exciting
physics and for further benchmarking between experimental and
computational molecular conduction.

We thank S. Datta, M. Lundstrom, T. Rakshit, D. Kienle and M.
Hersam for useful discussions. This project has been supported by
ARO-DURINT, DARPA, SRC and NCN.


\begin{thebibliography}{100}
\bibitem{rsc} We have used a simple model for self-consistency since
the tips are far from the molecule so that the voltage drops
mainly across the tip-sample gap and the molecular potential stays
tied to the substrate. Under these conditions, the transmission
peak energies directly translate into peak voltages with no
Laplace scaling factors. The strong substrate-molecule coupling
keeps the molecule out of the Coulomb Blockade regime, so that
charging is adiabatic and the HOMO-LUMO gap stays unaffected.
\bibitem{reta} The HOMO-LUMO gap here is larger, possibly due to a
voltage division factor that can be readily incorporated in our
model.
\end{thebibliography}
\end{document}